\documentclass[%
 reprint,
 amsmath,amssymb,
 aps,
 prd
]{revtex4-1}

\usepackage{graphicx}
\usepackage{dcolumn}
\usepackage{bm}
\usepackage{hyperref}
\usepackage{subfigure}
\usepackage{mathtools}

\begin{document}

\title{The scalarized Raychaudhuri identity and its applications}

\author{Eduard G. Mychelkin}
\email{mychelkin@aphi.kz}
\author{Maxim A. Makukov}%
\email{makukov@aphi.kz}
\affiliation{%
 Fesenkov Astrophysical Institute \\
 050020, Almaty, Republic of Kazakhstan 
}%

\date{\today}

	\begin{abstract}
		We show that the covariant Raychaudhuri identity describing kinematic characteristics of space-time admits a representation involving a geometrical scalar $\xi$ which, depending on circumstances, might be related to, e.g., relativistic temperature or cosmological scalar field. With an appropriately chosen spacetime deformation tensor (fixing the symmetry of a problem under consideration), such scalarization opens a wide scope for physical applications. We consider few such applications including dynamics of cosmological (anti)scalar background, non-variational deduction of the field equations, scalar and black-hole thermodynamics and the reshaping of the Einstein equations into the Klein-Gordon equation in thermodynamic Killing space.
	\end{abstract}

\maketitle

\section{Introduction}
Kinematical characteristics of timelike congruences $\{u^\mu\}$ are related to the Ricci tensor ${R}_{\mu\nu}$ via the fundamental result of the Riemannian geometry:
\begin{equation}
{{R}_{\alpha \beta }}{{u}^{\alpha }}{{u}^{\beta }} = -\dot{\theta } - \tfrac{1}{3}{\theta }^{2}  -  2{{\sigma }^{2}} + 2{{\omega }^{2}} + {\dot{u}^{\alpha}}_{\,\,\, ;\alpha}.
\label{EhlersTrad}
\end{equation}
Although this expression is often referred to as the Raychaudhuri equation (see, e.g., Hawking \& Ellis \cite{Hawking1973} and Wald \cite{Wald1984}), in a sense, this might be considered as an historical jargon, because in fact this expression is an \emph{identity} as it follows algebraically from the Ricci identity and is satisfied by any metric.  It is reducible to a proper \emph{equation} after  substitution of the Ricci tensor with the energy-momentum tensor in accord with the Einstein equations. In view of this, we will refer to the expression (\ref{EhlersTrad}) as the Raychaudhuri identity, while the result of the substitution of the Einstein equations for ${R}_{\alpha\beta}$ in (\ref{EhlersTrad}) will be referred to as the Raychaudhuri equation. It should also be noted that in  covariant form commonly used today, the identity (\ref{EhlersTrad}) had been obtained by Ehlers \cite{Ehlers1961}, and that was acknowledged by Raychaudhuri who referred to Ehlers' covariant result in \cite{De1968}. However, in non-covariant form specific to Friedmannian metrics the corresponding expression for ${R^0}_0$ was obtained earlier by Raychaudhuri \cite{Raychaudhuri1955}; another non-covariant form specific to stationary metrics was considered by Landau \& Lifshits \cite{Landau1994} (for additional historical notes, see, e.g., \cite{Kar2007}). 

In 2011, Abreu \& Visser \cite{Abreu2011} had obtained the generalization of (\ref{EhlersTrad}), replacing the unit congruence $u^\mu$ in this relation with an arbitrary (timelike, spacelike, or null) non-normalized vector field $\xi^{\mu}$ (see Eq. 49 in \cite{Abreu2011}; here we use our notation):
\begin{eqnarray}
	{R}_{\mu\nu}\xi^\mu \xi^\nu &=& \left( \xi^\alpha {\xi^\mu}_{;\alpha} \right)_{;\mu} - \left({\xi^\alpha}_{;\alpha} \xi^\mu \right)_{;\mu} + \left( {\xi^\alpha}_{;\alpha}  \right)^2 \nonumber \\
	&-& \xi_{(\alpha;\beta)}\xi^{(\alpha;\beta)} + \xi_{[\alpha;\beta]}\xi^{[\alpha;\beta]}.
\label{AV}
\end{eqnarray}
The quantity ${R}_{\mu\nu}\xi^\mu \xi^\nu$, however, seems to be not so straightforward from the standpoint of physical applicability, unlike the standard Raychaudhuri scalar ${R}_{\mu\nu}u^\mu u^\nu =  \xi^{-2}{R}_{\mu\nu}\xi^\mu \xi^\nu$. We aim at obtaining similar generalization of (\ref{EhlersTrad}), retaining the left side in the standard form and focusing on the most important case of timelike congruences (as for spacelike and null congruences, see \cite{Abreu2011} and \cite{Mychelkin2017a}). Unlike in \cite{Abreu2011}, our approach is based on the following two guiding principles.

First, we employ the symmetries of the non-normalized vector field $\xi^\mu$ in terms of the \emph{spacetime deformation tensor} defined as the Lie derivative of the metric tensor with respect to the given field $\xi^\mu$. 

The second principle is that from the vector field $\xi^{\mu} = \xi u^\mu$ we detach the magnitude, i.e. the scalar field $\xi$ which might be related to such physical variables as temperature, redshift, etc. Such \emph{scalarization} leads to what we call the $\xi$-formalism, opening a wide scope for physical applications some of which are illustrated in this paper.  E.g., Hawking \& Ellis \cite{Hawking1973} employed the similar principle to deduce the Einstein equations in a non-variational way (see also below).

\section{Scalarization of the Raychaudhuri identity}
\subsection{Spacetime deformation tensor}
From the definition of the Riemann curvature tensor ${{R}^{\rho}}_{\mu \nu \lambda }$ specified with respect to a unit timelike congruence ${u}^{\mu }$, ${{u}^{\alpha }}{{u}_{\alpha }}=1$,
\begin{equation}
{{u}_{\mu ;\nu \lambda }}-{{u}_{\mu ;\lambda \nu }}={{u}^{\alpha }}{{R}_{\alpha \mu \nu \lambda }},
\label{RiemannU}
\end{equation}
one obtains the Raychaudhuri identity in its primary minimal form, i.e. without the decomposition of the term ${{u}_{\alpha ;\beta }}{{u}^{\beta ;\alpha }} $ into irreducible parts \cite{Mychelkin2017a}:
\begin{equation}
\begin{aligned}
{{R}_{\alpha \beta }}{{u}^{\alpha }}{{u}^{\beta }} &={{u}^{\alpha }}_{;\beta \alpha }{{u}^{\beta }} - {u^\alpha}_{;\alpha\beta}u^\beta \\
	& = {u^\alpha}_{;\beta\alpha}u^\beta -  \dot{\theta} \\
	& =    {\dot{u}^{\alpha}}_{\,\,\, ;\alpha}   - {{u}_{\alpha ;\beta }}{{u}^{\beta ;\alpha }}   -     \dot{\theta}.
\label{EhlersPrimary}
\end{aligned}
\end{equation}
Here, $\theta = {{u}^{\alpha }}_{;\alpha }$ is expansion, and overdot denotes proper time derivative $\dot{(\,\,)} \equiv (\,\,)_{;\alpha} u^\alpha \equiv {{u}^{\alpha }}{{\nabla }_{\alpha }}$.

The relation (\ref{RiemannU}) is a particular case of a more general Ricci identity specified with respect to an arbitrary vector field ${\xi }^{\mu}$, the magnitude of which might be represented as a geometrical scalar field $\xi$:
\begin{equation*}
{{\xi }_{\mu ;\nu \lambda }}-{{\xi }_{\mu ;\lambda \nu }}={{\xi }^{\alpha }}{{R}_{\alpha \mu \nu \lambda }}, 
\end{equation*}
where in time-like case ${{\xi }^{\mu }} = \xi {{u}^{\mu }}$. Contraction over the indices $\mu$ and $\lambda$ yields:
\begin{equation}
	 {R}_{\alpha\nu}\xi^\alpha = \xi {R}_{\alpha\nu}u^\alpha = {\xi^\lambda}_{;\nu\lambda}-{\xi^\lambda}_{;\lambda\nu}. 
\label{RiemannXiContract}
\end{equation}
The right side of (\ref{RiemannXiContract}) might now be expressed through the so-called spacetime deformation tensor
\begin{equation}
 D_{\mu\nu} \coloneqq \mathcal{L_{\xi}}g_{\mu\nu}=\xi_{\mu;\nu}+\xi_{\nu;\mu}=2\xi u_{(\mu;\nu)} + 2\xi_{;(\mu}u_{\nu)},
\label{defT}
\end{equation}
defined as the Lie derivative of the metric tensor with respect to the given vector field ${{\xi }^{\mu }}$. The advantage of employing such tensor is that it provides a flexible tool to treat various symmetries of spacetime under consideration. From (\ref{defT}) it follows that:
\begin{equation}
{\xi^\lambda}_{;\nu\lambda} = {D^\lambda}_{\nu;\lambda} - \square \xi_\nu, \quad \square \xi_\nu \equiv {{\xi_\nu}^{;\lambda}}_{;\lambda},
\label{Dalamb}
\end{equation}
and
\begin{equation}
	{{D}^{\alpha }}_{\alpha }=2{{\xi }^{\alpha }}_{;\alpha } \quad \Rightarrow \quad  {\xi^\lambda}_{;\lambda\nu} = \tfrac{1}{2}{D^\lambda}_{\lambda;\nu}.
\label{traceD}
\end{equation}
Then, substituting (\ref{Dalamb}) and (\ref{traceD}) into (\ref{RiemannXiContract}), we get
\begin{equation} \xi{{R}_{\alpha \nu }}{{u }^{\alpha }} =  -\tfrac{1}{2}{{D}^{\alpha }}_{\alpha ;\nu }+{{D}^{\lambda }}_{\nu ;\lambda } -\square {{\xi }_{\nu }}={{f}_{\nu }}-\square({{\xi } u_{\nu }}),
\label{RicciXi}
\end{equation}
where the new object $f_\nu$ might be called the deformation vector:
\begin{equation}
	{{f}_{\nu }} \equiv {{{{D}^{\alpha }}_{\nu }}_{;\alpha }} -\tfrac{1}{2}{{{{D}^{\alpha }}_{\alpha }}_{;\nu}}.
\label{defV}
\end{equation}
Written in terms of the scalar $\xi$ and the unit congruence $u^\mu$, it has the following form:
\begin{eqnarray*}
	f_\mu &=& \xi_{;\mu\nu}u^\nu +  \xi {u^\nu}_{;\mu \nu}  + \xi^{;\nu}u_{\nu;\mu} + 2 \xi^{;\nu}u_{\mu;\nu}  \nonumber \\
	&&+  u_\mu \square \xi  + \xi \square u_\mu - \dot{\xi}_{;\mu} - \xi \theta_{;\mu},
\end{eqnarray*}
and thus
\begin{equation}
	f_\alpha u^\alpha = \xi u^\beta {u^\alpha}_{;\beta \alpha} + \square \xi + \xi u^\alpha \square u_\alpha - \xi \dot{\theta}.
\label{fmunu}
\end{equation}
So, we have derived basic relations of the $\xi$-formalism to start with.

\subsection{Primary form}
Now, projecting (\ref{RicciXi}) onto $u^\mu$-congruence again, and separating out the term $R_{\mu\nu} u^\mu u^\nu$, we obtain
\begin{eqnarray*}
	R_{\mu\nu} u^\mu u^\nu &=& \xi^{-1}f_\nu u^\nu - \xi^{-1} u^\nu \square(\xi u_\nu) \nonumber \\
	 &=& \xi^{-1}f_\nu u^\nu - \xi^{-1} \square \xi -  {{u}^{\nu }}\square {{u}_{\nu }}.
\end{eqnarray*}

\noindent It might be shown that
\begin{equation}
	{{u}^{\nu }}\square {{u}_{\nu }}= - {{u}_{\alpha ;\beta }}{{u}^{\alpha ;\beta }}\,\,,
\label{unudalam}
\end{equation}
 \noindent and so we ultimately have the following primary form of the scalarized Raychaudhuri identity:
 \begin{equation}
{{R}_{\alpha \beta }}{{u}^{\alpha }}{{u}^{\beta }} = -{{\xi }^{-1}}\square \xi +{{\xi }^{-1}}{{f}_{\alpha }}{{u}^{\alpha }}+{{u}_{\alpha ;\beta }}{{u}^{\alpha ;\beta }},
\label{EhlersG}
\end{equation}
As opposed to (\ref{EhlersPrimary}), this expression functionally relates the time-time projection of the Ricci tensor not only to the congruence $u^\mu$, but also to the geometrical scalar $\xi$ and, implicitly, to the deformation tensor ${D}_{\mu\nu}$ which, in general, might be chosen independently as a subsidiary condition.

\subsection{Decomposed form}
Next, we employ the decomposition 
 \begin{equation}
 {{u}_{\alpha ;\beta }}{{u}^{\beta ;\alpha }}={{u}_{\alpha ;\beta }}{{u}^{\alpha ;\beta }}-4{{\omega }^{2}}-{{\dot{u}}_{\alpha }}{{\dot{u}}^{\alpha }}
 \label{UabUab1}
 \end{equation}
following from the known expressions $${{u}_{\alpha ;\beta }}={{\omega }_{\alpha \beta }}+{{\sigma }_{\alpha \beta }}+\tfrac{1}{3}\theta {{h}_{\alpha \beta }}+{{\dot{u}}_{\alpha }}{{u}_{\beta }}$$
and
\begin{equation}
	{{u}_{\alpha ;\beta }}{{u}^{\beta ;\alpha }} =  \tfrac{1}{3}{\theta }^{2}  +  2{{\sigma }^{2}} - 2{{\omega }^{2}},
	\label{UabUba}
\end{equation}
with vorticity $\omega$ and shear $\sigma$ defined by
$$
\omega^2=\tfrac{1}{2}\omega_{\alpha \beta}\omega^{\alpha \beta}, \quad
\omega_{\mu \nu} = {h_\mu}^\alpha {h_\nu}^\beta u_{[\alpha;\beta]} 
$$
and
$$
\sigma^2=\tfrac{1}{2}\sigma_{\alpha \beta}\sigma^{\alpha \beta}, \quad
\sigma_{\mu \nu} = {h_\mu}^\alpha {h_\nu}^\beta u_{(\alpha;\beta)}-\frac{1}{3}\theta {{h}_{\mu \nu }},
$$
correspondingly. Here, $h_{\mu \nu} = g_{\mu \nu} - u_\mu u_\nu$ is the projector operator. Then, as is known, the traditional Raychaudhuri identity might be recast from its primary form (\ref{EhlersPrimary})  into decomposed form (\ref{EhlersTrad}). In the same manner, applying (\ref{UabUab1}) and (\ref{UabUba}), the scalarized  Raychaudhuri identity might be cast from the primary form (\ref{EhlersG}) into the decomposed form:
\begin{eqnarray}
{{R}_{\alpha \beta }}{{u}^{\alpha }}{{u}^{\beta }} = &-&{{\xi }^{-1}}\square \xi +{{\xi }^{-1}}{{f}_{\alpha }}{{u}^{\alpha }}+2{{\omega }^{2}} \nonumber \\
&+&{{\dot{u}}_{\alpha }}{{\dot{u}}^{\alpha }}+\tfrac{1}{3}{\theta }^{2}  +  2{{\sigma }^{2}}. 
\label{EhlersScDecomp}
\end{eqnarray}
Note that here the sign of the last two terms differs from that in (\ref{EhlersTrad}). This is due to the first two terms which are absent in (\ref{EhlersTrad}). After expressing the second term via the definition (\ref{fmunu}), and using (\ref{unudalam}), (\ref{UabUab1}) and (\ref{UabUba}), the formula (\ref{EhlersScDecomp}) might be reduced to (\ref{EhlersTrad}).

\section{Comparison with the Abreu-Visser result} 
\label{sec:AbreuVisserConnection}
Here we prove that our scalarized version of the Raychaudhuri identity (\ref{EhlersG}) is equivalent to the Abreu \& Visser version (\ref{AV}) with  a time-like non-normalized congruence. This might be shown by reducing both expressions to the same form. Thus, substituting (\ref{fmunu}) into (\ref{EhlersG}) and taking into account (\ref{unudalam}) one arrives at the standard (normalized) Raychaudhuri identity (\ref{EhlersPrimary}). Now, consider the Abreu-Visser  version and substitute $\xi^\mu = \xi u^\mu$. Then the first three terms on the right in (\ref{AV}) become, correspondingly:
\begin{eqnarray*}
\left( \xi^\alpha {\xi^\mu}_{;\alpha} \right)_{;\mu} &=& \dot{\xi}^2 + 3 \xi \xi_{;\alpha} \dot{u}^\alpha + \xi u^\alpha u^\beta \xi_{;\alpha \beta} + \xi \dot{\xi} \theta \\
 && \quad  + \,\, \xi^2 u_{\alpha;\beta} u^{\beta;\alpha} + \xi^2 u^\alpha {u^\beta}_{;\alpha \beta},\\
-\left( {\xi^\alpha}_{;\alpha} \xi^\mu  \right)_{;\mu} &=& -\dot{\xi}^2 - \xi \dot{\xi}_{;\alpha}u^\alpha - 3 \xi \dot{\xi} \theta - \xi^2 \theta^2 -\xi^2 \dot{\theta},\\
\left( {\xi^\alpha}_{;\alpha}  \right)^2 &=& \left( \dot{\xi} + \xi \theta \right)^2 = \dot{\xi}^2 + 2 \xi \dot{\xi}\theta + \xi^2\theta^2,
\end{eqnarray*}
while the last two terms in (\ref{AV}) together yield
$$
- \xi_{(\alpha;\beta)}\xi^{(\alpha;\beta)} + \xi_{[\alpha;\beta]}\xi^{[\alpha;\beta]} = -\dot{\xi}^2 - 2 \xi \xi_{;\alpha} \dot{u}^\alpha - \xi^2 u_{\alpha;\beta}u^{\beta;\alpha}.
$$
Summing up and taking into account that $\dot{\xi}_{;\alpha}u^\alpha = (\xi_{\beta}u^\beta)_{;\alpha}u^\alpha=\xi_{;\alpha \beta}u^\alpha u^\beta + \xi_{;\alpha}\dot{u}^\alpha$, we have instead of (\ref{AV}):
$$
\xi^2 {R}_{\mu\nu}u^\mu u^\nu =  \xi^2 u^\alpha {u^\beta}_{;\alpha \beta} - \xi^2 \dot{\theta},
$$
which is the same as (\ref{EhlersPrimary}), Q.E.D.

In fact, all versions of the Raychaudhuri identity -- normalized (standard), non-normalized and scalarized -- are mathematically equivalent because their left side is (or might be) represented as the same  Raychaudhury scalar ${R}_{\mu\nu}u^\mu u^\nu$, so that expressions on the right can always be transformed one into another. The essential difference is only in how certain symmetries of congruences and metrics under consideration enter functionally into each of these versions.

In particular, in the Killing case ${D}_{\mu \nu }=0$ and we have ${{f}^{\mu }}=0$, ${{u}_{\alpha ;\beta }}=-{{\dot{u}}_{\alpha }}{{u}_{\beta }}$ (here ${{\dot{u}}_{\alpha }}$ is four-acceleration), and so the scalarized Raychaudhuri identity (\ref{EhlersG}) reduces  to
 \begin{equation}
 {{R}_{\alpha \beta }}{{u}^{\alpha }}{{u}^{\beta }}=-{{\xi }^{-1}}\square \xi +{{u}_{\alpha ;\beta }}{{u}^{\alpha ;\beta }}.
 \label{KillingPrimaryEhlers}
 \end{equation}
 Abreu \& Visser had found analogous expression in the following form (with $\xi^{\mu}$ being the Killing vector):
 \begin{equation}
 	{R}_{\alpha\beta} \xi^\alpha \xi^\beta = \nabla_\alpha \left( \xi^\beta \nabla_\beta \xi^\alpha \right) + \xi_{[\alpha;\beta]} \xi^{[\alpha;\beta]}
 	\label{Abreu}
 \end{equation}
 (see Formula (52) in \cite{Abreu2011}). To prove equivalence of (\ref{KillingPrimaryEhlers}) and (\ref{Abreu}), it may be shown that the first term on the right side of (\ref{Abreu}) is reducible to
 \begin{equation*}
 \nabla_\alpha \left( \xi^\beta \nabla_\beta \xi^\alpha \right)=-\xi \square \xi - \xi_{;\alpha}\xi^{;\alpha},
 \end{equation*}
 while the second term is
 \begin{equation*}
 \begin{aligned}
 \xi_{[\alpha;\beta]} \xi^{[\alpha;\beta]} & = \xi_{\alpha;\beta} \xi^{[\alpha;\beta]}  = \tfrac{3}{2}\xi_{;\alpha}\xi^{;\alpha} + \xi^2 u_{\alpha;\beta}u^{[\alpha;\beta]} \\ 
 & = \tfrac{3}{2}\xi_{;\alpha}\xi^{;\alpha} + \xi^2 u_{\alpha;\beta}u^{\alpha;\beta} - \xi^2 u_{\alpha;\beta}u^{(\alpha;\beta)} \\
 & =\tfrac{3}{2}\xi_{;\alpha}\xi^{;\alpha} + \xi^2 u_{\alpha;\beta}u^{\alpha;\beta} - \tfrac{1}{2}\xi_{;\alpha}\xi^{;\alpha} \\
 & = \xi_{;\alpha}\xi^{;\alpha}+\xi^2 u_{\alpha;\beta}u^{\alpha;\beta}, \quad 
 \end{aligned}
 \end{equation*}
 and so their sum, taking into account that the left side of (\ref{Abreu}) is ${R}_{\alpha\beta} \xi^\alpha \xi^\beta = \xi^2R_{\alpha\beta} u^\alpha u^\beta$, yields (\ref{KillingPrimaryEhlers}). Q.E.D.

\section{Typical choices of the deformation tensor} 
\label{sec:symcases}

The Killing, four-conformal, space-conformal or time-conformal symmetries are defined as follows:
\begin{subequations}
\begin{eqnarray}
\text{Killing:} \,\, {D}_{\mu \nu } = {{\xi }_{\mu ;\nu }}+{{\xi }_{\nu ;\mu }} &=&  0,\qquad  \label{DOptions:a} \\
\text{four-conf.:} \,\, {{D}_{\mu \nu }} = {{\xi }_{\mu ;\nu }}+{{\xi }_{\nu ;\mu }} &=&  2\Phi(x^\lambda) {{g}_{\mu \nu }} , \qquad\label{DOptions:b} \\
\text{space-conf.:} \,\, {{D}_{\mu \nu }} = {{\xi }_{\mu ;\nu }}+{{\xi }_{\nu ;\mu }} &=&  2 \Psi(x^\lambda){{h}_{\mu \nu }} , \qquad\label{DOptions:c} \\
\text{time-conf.:} \,\, {{D}_{\mu \nu }} = {{\xi }_{\mu ;\nu }}+{{\xi }_{\nu ;\mu }} &=& 2 \Xi(x^\lambda)u_\mu u_\nu, \qquad \label{DOptions:d} 
\end{eqnarray}
\label{DOptions}
\end{subequations}
The four-conformal symmetry implies $\Phi=\Psi=\Xi$, since $2\Phi {g}_{\mu\nu} \equiv 2 \phi \left( {h}_{\mu\nu} + u_\mu u_\nu  \right) = 2\Phi {h}_{\mu\nu} + 2 \Phi u_\mu u_\nu$, provided that $\xi_\mu^{(b)} = \xi_\mu^{(c)} + \xi_\mu^{(d)}$, where the upper index indicates corresponding equation in (\ref{DOptions}).

The conformal factor $2 \Phi$ in (\ref{DOptions:b}) should not be confused with the  factor $\Omega^2$ in the conformal transformation ${\tilde{g}}_{\mu\nu} = \Omega^2 {g}_{\mu\nu}$. Substituting $\tilde{g}_{\mu\nu}$  and ${g}_{\mu\nu}$ into (\ref{DOptions:b}), it might be shown that these factors are related by (see, e.g., \cite{Svesko2018}):

\begin{equation}
\Phi = \frac{\xi^\alpha \nabla_\alpha \Omega }{\Omega} = \xi\frac{ \dot{\Omega}}{ \Omega}.
\label{confFactor}
\end{equation}
Each of the expressions in (\ref{DOptions}) might be included into the primary form of the Raychaudhuri identity (\ref{EhlersG}) with corresponding deformation vector $f_\alpha$ (\ref{defV}). As a result, we obtain four identities functionally differing only in  the last term on the right:
\begin{subequations}
\begin{eqnarray}
\label{EhlersKillingPrim}
{{R}_{\alpha \beta }}{{u}^{\alpha }}{{u}^{\beta }} &=& -{{\xi }^{-1}}\square \xi  +{{u}_{\alpha ;\beta }}{{u}^{\alpha ;\beta }},\label{EhlersKillingPrim:a}\\
{{R}_{\alpha \beta }}{{u}^{\alpha }}{{u}^{\beta }} &=&  -{{\xi }^{-1}}\square \xi  +{{u}_{\alpha ;\beta }}{{u}^{\alpha ;\beta }} - 2\xi^{-1}u^\alpha \nabla_\alpha{\Phi},\nonumber\\
{{R}_{\alpha \beta }}{{u}^{\alpha }}{{u}^{\beta }} &=& -{{\xi }^{-1}}\square \xi  +{{u}_{\alpha ;\beta }}{{u}^{\alpha ;\beta }} -\xi^{-1}( 2\Psi \theta +3u^\alpha\nabla_\alpha{\Psi}  ),  \nonumber \\
{{R}_{\alpha \beta }}{{u}^{\alpha }}{{u}^{\beta }} &=&  -{{\xi }^{-1}}\square \xi  +{{u}_{\alpha ;\beta }}{{u}^{\alpha ;\beta }} + \xi^{-1}( 2\Xi \theta + u^\alpha\nabla_\alpha{\Xi}  ). \nonumber
\end{eqnarray}
\label{fourPrimary}
\end{subequations}
So, for $\Xi\neq \Psi$ the space-conformal and time-conformal cases might be considered as independent. The idea of selective action of conformal transformations on submanifolds of a Riemannian manifold had been evolved in a number of works -- see, e.g., \cite{Tanno1965} and, in form-covariant manner, \cite{Mychelkin1991}.  The peculiarity of the cases (c)  and (d) is that the conformal weight of physical quantities under consideration is defined by the power of length dimensionality for (c) and of time dimensionality for (d).

In some cases, apart from (\ref{DOptions:b}), one might consider the conformal optical metric (see, e.g., \cite{Ehlers1967Optik}) in the definition of the deformation tensor:
\begin{equation*}
	\text{optical-conformal:} \quad {{D}_{\mu \nu }} = {{\xi }_{\mu ;\nu }}+{{\xi }_{\nu ;\mu }} =  2\bar{\Phi}(x^\lambda) {{\bar{g}}_{\mu \nu }},
\end{equation*}
with ${\bar{g}}_{\mu\nu} = {g}_{\mu\nu} - \left( 1 - n^{-2}  \right) u_\mu u_\nu$, where $n$ is the effective refraction index of the medium under consideration. Such approach had been developed with application to systems with spontaneous creation/annihilation of particles \cite{Zimdahl2000}, which also might lead to the symmetry of the type (\ref{DOptions:c}), as shown in \cite{Zimdahl1998}.

\section{Cosmological applications} 
\label{sec:cosmological_application}
We seek for the simplest natural prolongation of the minimal antiscalar field (which is well-justified by observations; for details on antiscalarity see \cite{Makukov2018}) to massive field at cosmological scales. To this end, consider the energy-momentum tensor of a cosmological scalar field with non-zero mass-term and negative $\Lambda$-term:
\begin{equation*}
\begin{aligned}
&T_{\mu \nu }^{sc} - \frac{1}{8\pi }\Lambda{g}_{\mu \nu } = \\
=\frac{1}{4\pi }&\left[ {{\phi }_{,\mu }}{{\phi }_{,\nu }}-\frac{1}{2}{{g}_{\mu \nu }} \left( {{\phi }_{,\alpha }}{{\phi }^{,\alpha }} - {m^{2}}{{\phi }^{2}}\right) \right] - \frac{1}{8\pi }\Lambda{g}_{\mu \nu }.
\end{aligned}	
\end{equation*}
The choice of the negative sign for the cosmological term is justified below.
Then, from the field equations in scalar (upper signs) and antiscalar (lower signs) regimes, ${G}_{\mu\nu}=\pm 8 \pi {T}_{\mu\nu}^{sc} - \Lambda {g}_{\mu\nu} $, we obtain:
\begin{equation*}
{R}_{\mu\nu} = \pm 2 \phi_\mu \phi_\nu \mp  m^2 \phi^2 {g}_{\mu\nu} + \Lambda {g}_{\mu\nu},
\end{equation*}
and so it is necessary to satisfy the system represented by the Raychaudhuri equation,
\begin{equation}
{R}_{\mu\nu}u^\mu u^\nu = \pm 2 \dot{\phi}^2 \mp m^2 \phi^2 + \Lambda,
\label{EinASFfull}
\end{equation}
and the Klein-Gordon equation,
\begin{equation}
\square\phi + m^2 \phi=   \frac{1}{\sqrt{-g}}\frac{\partial \left( \sqrt{-g} g^{\mu \alpha}\partial_\alpha \phi\right)}{\partial x^\mu} + m^2 \phi= 0.
\label{KG}
\end{equation}
In general, solving the system (\ref{EinASFfull})-(\ref{KG}) is rather difficult. However, there exist  important cosmological solutions in the Friedmanian-type metrics
\begin{equation}
	ds^2 = dt^2 - e^{F(t)}\left( dr^2 + r^2 d \Omega^2  \right),
	\label{metricOriginal}
\end{equation}
which we consider next.

\subsection{Scalar solution}
We solve directly the Einstein-Klein-Gordon system in scalar regime with negative $\Lambda$-term, i.e. for ${G}_{\mu\nu} = 8\pi {T}_{\mu\nu}^{sc} - \Lambda {g}_{\mu\nu}$ together with (\ref{KG}). Then, writing the equations for ${G_0}^0$ and ${G_1}^1$ in the metric (\ref{metricOriginal}) we get, correspondingly:
\begin{equation}
	\frac{3}{4}\dot{F}^2 = \dot{\phi}^2 + m^2 \phi^2 - \Lambda 
	\label{fff1}
\end{equation}
and
\begin{equation}
	\frac{3}{4}\dot{F}^2 + \ddot{F} = - \dot{\phi}^2 + m^2 \phi^2 - \Lambda .
	\label{fff2}
\end{equation}
The difference  of the two is $\ddot{F}=-2 \dot{\phi}^2$. Adopting the Papapetrou ansatz that metric dependence on coordinates enters only through $\phi$, i.e. ${g}_{\mu\nu}(x^\alpha) = {g}_{\mu\nu}(\phi(x^\alpha))$ \cite{Papapetrou1954b}, we seek for  solutions in the following form:
\begin{equation}
F(t)= - \phi^2(t),
\label{ansatzF}
\end{equation}
from which we get $\dot{F}= -2 \phi \dot{\phi}$ and $\ddot{F} = - 2 \dot{\phi}^2  - 2 \phi \ddot{\phi}$. To avoid contradiction with (\ref{fff1}) and (\ref{fff2}) one should pose $\ddot{\phi}=0$, implying that $\phi$ is a linear function of $t$. Then the system (\ref{fff1}), (\ref{fff2}) reduces to a single equation,
\begin{equation}
	\dot{\phi}^2 =  \frac{m^2 \left( \phi^2 - \frac{\Lambda}{m^2}\right) }{3\left(\phi^2 - \frac{1}{3}\right)}.
\end{equation}
It might be shown that in general this equation might be solved in terms of elliptic integrals. However, the required linearity of $\phi(t)$ might be satisfied only  under the following integrability condition:
\begin{equation}
	\Lambda = \frac{m^2}{3},
	\label{Lam}
\end{equation}
yielding (up to the sign) 
\begin{equation}
	\phi = \frac{m}{\sqrt{3}} \left( t - t_0  \right) = \sqrt{\Lambda} \left( t - t_0  \right).
	\label{phi}
\end{equation}

The resulting square of mass in (\ref{Lam}) is found to be $m^2 = 3\Lambda \approx 3\times 10^{-56}$ $\text{cm}^{-2}$ or $m \approx 10^{-33}$ $\text{eV}$. The real (positive) character of the relation (\ref{Lam}) corresponds to our choice of the negative sign for the $\Lambda$-term (cf. \cite{Maciejewski2008JPA}, where the positive $\Lambda$-term yields negative square of mass).

Thus, (\ref{metricOriginal}) proves to be the Gaussian metric 
\begin{equation}
	ds^2=  dt^2 - e^{-\Lambda(t-t_0)^2}\left( dr^2 + r^2 d \Omega^2  \right)
	\label{gauss}
\end{equation}
which may be considered as a fundamental cosmological solution appropriate for the description of dark energy background \cite{Mychelkin2015}. 

Finally, we check the compatibility of the Einstein equations with the Klein-Gordon equation in the metric (\ref{metricOriginal}) under the subsidiary condition $\ddot{\phi}=0$:
\begin{equation*}
\square\phi + m^2 \phi = \frac{3}{2} \dot{F} \dot{\phi} + m^2 \phi = 0 .
\end{equation*}
This yields $3\dot{\phi}^2 = m^2$, as it should according to (\ref{phi}).

\subsection{Antiscalar solution}
Analogous calculation for  antiscalar case, ${G}_{\mu\nu} = -8\pi {T}_{\mu\nu}^{sc} - \Lambda {g}_{\mu\nu}$, implying the replacement $\phi \rightarrow i \phi$ \cite{Makukov2018} and so leading to the ansatz
\begin{equation}
	F=\phi^2
\label{ansatzFASF}	
\end{equation}
instead of (\ref{ansatzF}), yields the same solution (\ref{gauss}) under the condition $\ddot{\phi} = 0$, as well as compatibility with the corresponding Klein-Gordon equation.

Alternatively, working directly with the Raychaudhuri equation and using the metric (\ref{metricOriginal}) with antiscalar ansatz (\ref{ansatzFASF}) in comoving frame ($u^\mu = \delta^\mu_0/\sqrt{g_{00}}$) we obtain:
\begin{equation}
	{R}_{\mu\nu}u^\mu u^\nu = -3 \left( \dot{\phi}^2 + \phi^2 \dot{\phi}^2 + \phi \ddot{\phi} \right).
\end{equation}
Substituting this into (\ref{EinASFfull}) and taking into account that in this case again $\ddot{\phi} = 0$, we finally get:
\begin{equation*}
	\dot{\phi}^2 + 3 \dot{\phi}^2 \phi^2 + m^2 \phi^2  + \Lambda  = 0,
\end{equation*}
which might be rewritten as
\begin{equation*}
	\dot{\phi}^2 = - \frac{m^2 \left( \phi^2 + \frac{\Lambda}{m^2}\right) }{3\left(\phi^2 + \frac{1}{3}\right)}.
\end{equation*}
Under the conditions (\ref{Lam}) and (\ref{ansatzFASF}), this yields the solution with antiscalar (effectively imaginary) potential
\begin{equation}
	\phi = i\frac{m}{\sqrt{3}} \left( t - t_0  \right) = i\sqrt{\Lambda} \left( t - t_0  \right),
	\label{phiASF}
\end{equation}
and, correspondingly, again the Gaussian metric (\ref{gauss}). As it should be, the antiscalar Klein-Gordon equation  in the metric (\ref{metricOriginal}) with the condition (\ref{ansatzFASF}),
\begin{equation*}
 \left( \square + m^2  \right)\phi = \left( 3 \dot{\phi}^2 + m^2 \right) \phi = 0,
\end{equation*}
is identically satisfied with the solution (\ref{phiASF}).

Thus, both in scalar and antiscalar regimes with negative $\Lambda$-term we get the same cosmological  solution (\ref{gauss}) with  real-valued integrability condition (\ref{Lam}). In a sense, such situation is opposite to the $\Lambda$-vacuum case where de Sitter and anti-de Sitter metrics represent topologically different solutions. However, in static case only antiscalar regime admits the solution (the Papapetrou metric) which proves to be in excellent agreement with observational data (see \cite{Makukov2018} and Sec.~(\ref{sec:HEdeductionEE}) below).

The significance of the two examples presented above is that due to the existence of such background (which might be identified with dark energy \cite{Mychelkin2015}) the values of the $\Lambda$-term and of mass of (anti)scalar field mediators appear to be coupled according to (\ref{Lam}). Moreover, in cosmological general relativity the mass- and $\Lambda$-terms, being of the same order of magnitude, should be included into (or excluded from) the (anti)scalar  stress-energy tensor only simultaneously.

\subsection{Cosmological conformal symmetry}
Now, we consider the obtained solution in terms of the spacetime deformation tensor. We begin with the general cosmological metric (\ref{metricOriginal}). 

Rewriting the deformation tensor (\ref{DOptions:b}) with partial derivatives as $$	{D}_{\mu\nu} = \mathcal{L}_{\xi} {g}_{\mu\nu} = \xi^\alpha \partial_\alpha {g}_{\mu\nu} + {g}_{\mu \alpha} \partial_\nu \xi^\alpha +{g}_{\nu \alpha} \partial_\mu \xi^\alpha = 2 \Phi {g}_{\mu\nu},$$
where $\xi^\mu = \xi u^\mu$, with $u^\mu = \delta^\mu_0/\sqrt{{g}_{00}}$ (comoving frames), and ${g}_{\mu\nu}$ represents Friedmannian metric (\ref{metricOriginal}) with scale factor $e^{F(t)}=a^2(t)$, we get
\begin{equation}
	D_{00} = 2 \dot{\xi}  = 2 \Phi  \quad \Rightarrow \quad \Phi = \dot{\xi}
	\label{e62}
\end{equation}
and then
\begin{eqnarray*}
	D_{11} &=&  - \xi u^\alpha \partial_\alpha (a^2) = -2 \Phi a^2 \quad \Rightarrow \\ 
	  \frac{\dot{\xi}}{\xi} &=& \frac{\dot{a}}{a} = H(t), \quad \text{or} \quad a = \frac{\xi}{\xi_0},
\end{eqnarray*}
where $\xi_0$ is the integration constant and $H(t)$ is the Hubble parameter. So, the expression (\ref{e62}) might be rewritten as 
\begin{equation*}
	\Phi =  \xi\frac{\dot{a}}{a} = \xi H.
\end{equation*}
Juxtaposing this with the general relation (\ref{confFactor}) we see that the conformal factor transforming the Minkowski spacetime into the Friedmannian one is $\Omega = a$, as it should be.
As a result, relation (\ref{DOptions:b})  becomes
\begin{equation*}
	\xi_{(\mu;\nu)}=\dot{\xi} {g}_{\mu\nu} = 2 \frac{\dot{a}}{a} \xi {g}_{\mu\nu},
\end{equation*}
and at the same time, the Friedmannian metric (\ref{metricOriginal}) acquires the following $\xi$-form:
\begin{equation*}
	ds^2 = dt^2-{\xi}^2(t)\left( dr^2 + r^2 d \Omega^2  \right).
\end{equation*}
For the Gaussian metric (\ref{gauss}) we, evidently, have $\xi = e^{-{\Lambda(t-t_0)^2}/{2}}$.

As for  conformal-type symmetries (\ref{DOptions:c}) and (\ref{DOptions:d}), they prove to be incompatible with Friedmannian metrics, as follows from  direct calculations. In a more general situation when $\xi = \xi(t, \mathbf{r})$ and $g_{00} = g_{00}(t, \mathbf{r})$ (e.g. for the bounded systems imbedded into cosmological background) the symmetries of the types (\ref{DOptions:c}) and (\ref{DOptions:d}) might appear relevant.

\section{The Killing symmetries}
\label{sec:HE}

For a static scalar field, the Killing symmetries (\ref{EhlersKillingPrim:a}) in time-like case imply $\dot{\phi }=0$. However, unlike in cosmology, finding the solutions of (\ref{EinASFfull})-(\ref{KG}) with non-zero mass-term represents a  non-trivial task. Meanwhile, for static (equilibrium) systems with minimal scalar field there exist a number of interesting problems which are solvable and might be naturally described in terms of the $\xi$-formalism being developed.

\subsection{Non-variational Hawking-Ellis deduction of the Einstein equations}
\label{sec:HEdeductionEE}
The presented $\xi$-formalism is intricately related to the Hawking-Ellis non-variational approach to the deduction of the Einstein equations. As a direct consequence of the scalarized Raychaudhuri identity  (\ref{KillingPrimaryEhlers}), we get:
\begin{eqnarray*}
{{R}_{\alpha \beta }}{{u}^{\alpha }}{{u}^{\beta }}&=&-{{\xi }^{-1}}\square \xi +{{u}_{\alpha ;\beta }}{{u}^{\alpha ;\beta }}\nonumber \\
&=&-{{\xi }^{-1}}\square \xi + {{\xi }^{-2}}\xi_{;\alpha}\xi^{;\alpha} \nonumber \\ 
&=&-{{\xi }^{-1}}{{\xi }_{;\alpha \beta }}({{g}^{\alpha \beta }}-{{u}^{\alpha }}{{u}^{\beta }}) \nonumber \\
&=&-{{\xi }^{-1}}{{h}^{\alpha \beta }}{{\xi }_{;\alpha \beta }},
\end{eqnarray*} 
Here we have made use of the fact that ${{u}_{\alpha ;\beta }}{{u}^{\alpha ;\beta }}   = {{\dot{u}}_{\alpha }}{{\dot{u}}^{\alpha }}$ by virtue of (\ref{UabUab1}) and (\ref{UabUba}) for $\theta = \omega = \sigma = 0$, with ${{\dot{u}}_{\alpha }}{{\dot{u}}^{\alpha }} = {{\xi }^{-2}}\xi_{;\alpha}\xi^{;\alpha}$ due to $\dot{\xi}^{\alpha}=-\xi^{;\alpha}$ (true for any Killing vector $\xi^\mu$), and also that
\begin{equation}
{{\xi }^{-2}}\xi_{;\alpha}\xi^{;\alpha} = {{\xi }^{-1}}{{\xi }_{;\alpha \beta }}{{u}^{\alpha }}{{u}^{\beta }},
\label{xitmp}
\end{equation}
as might be obtained  from the condition ${\ddot{\xi}}=0$. Such scalarized representation of the Raychaudhuri identity in final form 
\begin{equation}
{{R}_{\alpha \beta }}{{u}^{\alpha }}{{u}^{\beta }}=-{{\xi }^{-1}}{{h}^{\alpha \beta }}{{\xi }_{;\alpha \beta }} 
\label{toHW}
\end{equation}
was first derived by Hawking \& Ellis \cite{Hawking1973}, who also showed that in weak-field approximation ($\xi = \sqrt{{g}_{00}} \approx 1 - \phi$, $\phi \ll 1$) the right side reduces to the Laplacian of the Newtonian potential (which we choose positively defined). Then, in particular, the Poisson equation of the Newtonian gravity $\Delta \phi =-4\pi G\rho $ immediately follows after equating the Raychaudhuri scalar ${{R}_{\mu \nu }}{{u}^{\mu }}{{u}^{\nu }}$ to matter density $4\pi G\rho$.

On the other hand, since in relativity $\rho$ generalizes  to ${T}_{\mu\nu}u^\mu u^\nu$, this serves as an indication that there should exist a linear connection between the Ricci tensor and the energy-momentum tensor ${{T}_{\mu \nu }}$. This connection becomes unique after imposing the  divergence-free condition ${{T}^{\mu \nu }}_{;\nu }=0$. Then, applying the contracted Bianchi identity, one arrives at the field equations ${{R}_{\mu \nu }}-\tfrac{1}{2}R{{g}_{\mu \nu }}=\varkappa {{T}_{\mu \nu }}$ with $\varkappa = 8\pi G/c^4$, or simply $\varkappa = 8\pi$ if $G=c=1$.  Thus, the Einstein equations follow non-variationally from the scalarized Raychaudhuri identity with the Killing symmetry in weak-field approximation.

The Hawking-Ellis form (\ref{toHW}) of the scalarized Raychaudhuri identity is satisfied by any metric obeying a timelike Killing symmetry. As a special application,  we envisage the static spherically symmetric Papapetrou metric \cite{Papapetrou1954b}
\begin{eqnarray}
ds^2 &=& e^{-2\phi(r)}dt^2 - e^{2\phi(r)}(dr^2 +r^2 d \Omega^2) \nonumber\\
&=& \xi^2 dt^2 - \xi^{-2} (dr^2 +r^2 d \Omega^2),
\label{genstat}	
\end{eqnarray}
where scalar potential and the Killing vector magnitude are related as $\xi = \sqrt{{g}_{00}}={{e}^{-\phi }}$, and so, substituting (\ref{genstat}) into right side of (\ref{toHW}) we obtain:
\begin{eqnarray}
{{R}_{\mu \nu }}{{u}^{\mu }}{{u}^{\nu }}
&=& -{{e}^{\varphi }}\left( {{g}^{\mu \nu }}{{\nabla }_{\mu }}{{\nabla }_{\nu }}-{{u}^{\mu }}{{u}^{\nu }}{{\nabla }_{\mu }}{{\nabla }_{\nu }} \right){{e}^{-\varphi }}\nonumber\\
&=&{\phi^{;\alpha}}_{;\alpha} =  \Delta \phi,
\label{lapphi}
\end{eqnarray}
where we have taken into account that $\phi_{;\alpha}\phi^{;\alpha} = -\phi_{;\alpha \beta}u^\alpha u^\beta$ as a consequence of (\ref{xitmp}), and, at the last step, that $\dot{\phi}=0$. But unlike the weak-field approximation in the Hawking-Ellis procedure, the result (\ref{lapphi}) is \emph{not} an approximate but \emph{exact} expression for $\Delta \phi \equiv \phi_{;ij}g^{ij} \,\, (i,j = 1,2,3)$ in the metric (\ref{genstat}).

 In its turn, this metric represents the exact solution of the Einstein equations
\begin{equation}
{R}_{\mu\nu} = -8\pi \left(  {T}^{sc}_{\mu\nu} - \tfrac{1}{2}{g}_{\mu\nu}T^{sc} \right),
\label{EinASF}
\end{equation}
for the minimal scalar field energy-momentum tensor,
$$
{T}^{sc}_{\mu\nu} = \tfrac{1}{4 \pi}\left( \phi_{;\mu}\phi_{;\nu} - \tfrac{1}{2}{g}_{\mu\nu} \phi^{;\alpha}\phi_{;\alpha}  \right),
$$
taken here in antiscalar regime ${T}^{sc}_{\mu\nu}\, \rightarrow \, -{T}^{sc}_{\mu\nu}$ \cite{Mychelkin2015,Makukov2018}. Rewriting (\ref{EinASF}) as
${R}_{\mu\nu} = -2 \phi_\mu \phi_\nu$ we get ${R}_{\mu\nu}u^\mu u^\nu=-2 \dot{\phi}^2=0$, where the last equality is due to timelike Killing symmetry. Thus, (\ref{lapphi}) leads  to the Laplace equation $\Delta \phi=0$ with the solution $\phi = GM/r$ being the Newtonian potential entering  (\ref{genstat}).

At the same time,  the Raychaudhuri identity (\ref{toHW}) in case of  antiscalar background  for the metric (\ref{genstat}) turns into zero. In a sense, this result  explains why all observational effects predicted by antiscalar solutions prove to be so close to their vacuum analogues \cite{Makukov2018}. On the other hand, the actual value of (\ref{toHW}) is in fact irrelevant; for the result obtained only functional dependence on $\xi$ on the right side of (\ref{toHW}) matters, as was already noted above (see section \ref{sec:AbreuVisserConnection}).

\subsection{Antiscalar thermodynamics and BHs} 
 \label{sec:section_name}
Both in relativistic kinetics \cite{Synge1960} and thermodynamics the magnitude of time-like Killing vector $\xi^\mu = \xi u^\mu$ (dimensional in this case) represents the reciprocal temperature $\xi = 1/(kT) = \Theta^{-1}$. Then the local equilibrium solution of the functional Boltzmann equation being the standard covariant J\"{u}ttner-type distribution function (with $p_\mu p^\mu = m^2 c^2$),
 \begin{equation*}
 	f(x^\nu,p^\nu) =F(x^\nu) e^{ - p_\mu \xi^{\mu}c},
 \end{equation*}
produces the chain of moments which, in terms of the kinetic $\xi$-formalism, are written as:
\begin{eqnarray}
f^{(0)}(\xi) &=&  \int f(x^\mu, p^\mu) d \omega \equiv \Phi, \nonumber\\
f^{(1)}(\xi) &=& -\frac{\partial}{\partial \xi^\mu}f^{(0)} =  c \int p_\mu f d \omega = j_\mu, \label{xiForm}\\ 
f^{(2)}(\xi) &=&-\frac{\partial}{\partial \xi^\nu}f^{(1)} = c^2 \int p_\mu p_\nu f d \omega = c T_{\mu \nu},\nonumber
\end{eqnarray}
etc. Here, $j_\mu$ and $T_{\mu \nu}$ are the number flux density and the energy-momentum tensor, correspondingly, and $d\omega=\sqrt{-g}d^3p/p_0$ is an element of integration over momentum space. Integration can be performed in terms of the modified Bessel (Macdonald) functions $K_n(x)$ (see, e.g., \cite{Synge1960}):
\begin{eqnarray}
\Phi &=& 4 \pi m  F \frac{ K_1(m c^2 \xi)}{\xi},\nonumber\\
j^\mu &=& 4 \pi m^2 c^2 F \frac{ \xi^\mu K_2(m c^2 \xi)}{\xi^2},\label{macd}\\
T^{\mu \nu} &=& 4 \pi m^3 c^3 F\left[ \frac{ \xi^\mu \xi^\nu K_3(m c^2 \xi)}{\xi^3} - \frac{{g}^{\mu\nu}K_2(m c^2 \xi)}{m c ^2 \xi^2} \right],\nonumber
\end{eqnarray}
etc., to be normalized by $F = {n \xi}/(4 \pi m^2 c K_2(m c^2 \xi))$ with the number density $n = c^{-1} j_\alpha u^\alpha = c^{-1} {j_\alpha \xi^\alpha}/{\xi}$.

From (\ref{xiForm}) with $j^\mu = ncu^\mu$ we get  
\begin{equation*}
	{T}_{\mu\nu}= - \frac{\partial j_\mu}{c\partial \xi^\nu} = - u_\nu \frac{\partial \left( n {\xi_\mu}/{\xi}\right)}{\partial \xi}= -\frac{\partial n}{\partial \xi} u_\mu u_\nu - \frac{n}{\xi} {h}_{\mu\nu},
\end{equation*}
from which it follows for energy density and pressure:
\begin{equation}
\varepsilon = - \frac{\partial n}{\partial \xi}, \qquad p = \frac{n}{\xi}.
\label{grouphase}
\end{equation}
This might be verified with (\ref{macd}) in terms of Bessel functions as well.
Now, applying the relations (\ref{grouphase}) to the Gibbs equation (with $s$ being the entropy density and
$q$ the heat flux density), 
\begin{equation*}
dq = \xi^{-1} d\left(\frac{s}{n}\right) = d\left(\frac{\varepsilon}{n}\right) + pd\left(\frac{1}{n}\right)  = 0,
\end{equation*}
we obtain a differential equation with respect to $n=n(\xi)$:
$$
nn'' + \frac{1}{\xi}nn' -n'^2 = 0.
$$
Its first integral  represents the barotropic equation of state: 
\begin{equation*}
-w \frac{\partial n}{\partial \xi} = \frac{n}{\xi} \quad \Longleftrightarrow \quad p=w \varepsilon,
\end{equation*}
where $w$ is a constant, and finally we get:
\begin{equation}
n = C \xi^{-\frac{1}{w}}, \quad \varepsilon =  \frac{C}{w}\xi^{-\left(1+\frac{1}{w}\right)}, \quad  p = C\xi^{-\left(1+\frac{1}{w}\right)},
\label{nepsp}
\end{equation}
\begin{equation}
\frac{s}{k} = \frac{dp}{d\Theta} = - \frac{1}{\xi^2} \frac{dp}{d\xi} =  C\left(1+\frac{1}{w}\right)\xi^{-\tfrac{1}{w}} = \xi(\varepsilon+p),
\label{s}
\end{equation}	
where the chemical potential is taken to be zero, and $C=C(w)$ is the positive integration constant with physical dimensionality dictated by the value of $w$. 

The Einstein equations for perfect fluid with the equation of state $p=\varepsilon$, i.e. $w=  1$, can mimic minimal antiscalar field \cite{Makukov2018} and in static (equilibrium) case are satisfied by the Papapetrou metric (\ref{genstat}). Then the trace of the Einstein equations in terms of the $\xi$-formalized quantities in (\ref{nepsp}) is:
\begin{equation*}
-R = \varkappa \left(\varepsilon -3p\right) = \varkappa C  \frac{1-3w}{w}\xi_0^{-\left(1+\frac{1}{w}\right)},
\end{equation*}
i.e., for $w=1$,
\begin{equation}
R = 2 \varkappa C \xi_0^{-2},
\label{condT}
\end{equation}
where the special-relativistic invariant $\xi = 1/\Theta$ \, is replaced with the general-relativistic one $\xi_0 = 1/\Theta_0$ in conformity with Tolman's relation $\Theta_0 = \sqrt{g_{00}}\Theta$.

On the other hand, the  Ricci scalar might be calculated directly from the Papapetrou metric (\ref{genstat}) as
\begin{equation}
R = 2\frac{ G^2 M^2}{c^4 r^4} \exp\left(-\frac{2GM}{c^2r}\right) = 2\frac{ G^2 M^2}{c^4 r^4}g_{00}.
\label{condPapa}
\end{equation}
Equating (\ref{condT}) and (\ref{condPapa}), the local temperature  of antiscalar background is found to be 
\begin{equation}
\xi^{-1} = \Theta(r)   =\frac{1}{2\sqrt{2 \pi}} \sqrt{ \frac{G}{C} } \frac{M}{r^2}.
\label{speciallyForEd}
\end{equation}
Applying the general expression (\ref{speciallyForEd}) to the equipotential surface with $r=r_g=2GM/c^2$ we get for the value of local temperature $\Theta$ at this surface:
\begin{equation}
\Theta(r_g) = \frac{c^4}{8\sqrt{2 \pi}\sqrt{CG^3}} \frac{1}{M},
\label{BHtemp}
\end{equation}
which is similar to the Hawking black hole temperature:
\begin{equation}
\Theta_{\text{BH}} = k T_{\text{BH}} = \frac{\hbar c^3}{8\pi G}\frac{1}{M}.
\label{tbh}
\end{equation}
Comparison of (\ref{BHtemp}) with (\ref{tbh}) yields the corresponding value for $C$:
\begin{equation}
C = C(w=1) = \frac{\pi c^2}{2\hbar^2 G}.
\label{Cw1}
\end{equation}

It is known that the entropy of a black hole is proportional to the area of the event horizon, i.e., in Planck units, $S_{\text{BH}} = A/4$. To ascertain that the found value of $C$ is physically relevant one can compute the  entropy $S(r_g)$ within the domain $r=r_g$, taking into account (\ref{Cw1}). According to (\ref{s}), for $w=1$ we have $s = 2 k C \xi^{-1}$, and so
\begin{eqnarray}
S(r_g) &=& \int{s_\mu dV^\mu} = 4 \pi \int_0^{r_g}{s(r)r^2dr} \nonumber\\
&=&8 \pi C k \int_0^{r_g}{\xi^{-1}(r)r^2 dr} = k\frac{4 \pi  G }{\hbar c} M^2,\nonumber
\end{eqnarray}
where $s_\mu = su_\mu$ and $dV^\mu = u^\mu d^3V$. The last result coincides exactly with the mentioned value of black hole entropy expressed in standard units:
\begin{equation*}
	S_\text{BH} = k \frac{A}{4\ell_P^2}= k \frac{\pi r_g^2}{\ell_P^2} = k\frac{4 \pi  G }{\hbar c} M^2_{\text{BH}},
\end{equation*}
with $A$ being the area of the horizon and $\ell_P = \sqrt{ \hbar G/c^3}$ the Planck length. So, the antiscalar thermodynamics includes the relations of traditional black hole thermodynamics as a particular case for  $r=r_g$.

The $\xi$-formalism applied to thermodynamics of antiscalar background leads, in general, to a simpler interpretation of thermodynamic quantities as compared to the vacuum case. Here, the gravitational radius $r_g$ is not singled  out because  all equipotential surfaces for the Papapetrou spacetime are on equal footing, thereby filling up (analytically) all the space, and thus also  making the application of the holographic principle quite natural.

\subsection{Einstein equations in thermodynamic Killing space} 
\label{sub:einstein_equations_as_thermal_klein_gordon_equation}
When Jacobson \cite{Jacobson1995} had found the relation between the Einstein equations and thermodynamics he built upon the Raychaudhuri equation for null geodesic congruence, i.e., in fact, employed the conformal symmetry. Here, we search for an analogous relation in case of the timelike Killing symmetry.

To this end, starting from the kinetic  Einstein equations 
\begin{equation}
	{R}_{\mu\nu} - \frac{1}{2}R {g}_{\mu\nu} = \varkappa c \int p_\mu p_\nu f d \omega
	\label{kineins}
\end{equation}
and contracting them,
\begin{equation*}
	-R = \varkappa m^2 c^3 \int f d \omega \equiv \varkappa m^2 c^3 \Phi,
\end{equation*}
we find that the generating function $\Phi(\xi)$ is equal, up to a constant factor, to the Ricci scalar taken with the negative sign. Then, functional relations of the kinetic $\xi$-formalism (\ref{xiForm}) might be geometrized as follows:
\begin{equation}
\begin{aligned}
\Phi &= - \frac{1}{\varkappa m^2 c^3}R,\\
j_\mu &= \frac{1}{\varkappa m^2 c^3} \frac{\partial R}{\partial \xi^\mu},\\
T_{\mu \nu} &= - \frac{1}{\varkappa m^2 c^3} \frac{\partial^2R}{\partial \xi^\mu \partial \xi^\nu}, 
\end{aligned}
\label{xieins}
\end{equation}
etc. As a result, from the last relation in (\ref{xieins}) the Einstein equations (\ref{kineins}) become
\begin{equation}
	-\frac{1}{m^2 c^4}\frac{\partial^2R}{\partial \xi^\mu \partial \xi^\nu} = {R}_{\mu\nu} - \frac{1}{2}R {g}_{\mu\nu},
	\label{einsxi}
\end{equation}
with $\xi^\mu = \xi u^\mu$, \,\,\,${{\xi }_{\mu ;\nu }}+{{\xi }_{\nu ;\mu }}=0$, see (\ref{DOptions:a}). Transforming the left side of this expression,
\begin{equation*}
	\frac{\partial^2R}{\partial \xi^\mu \partial \xi^\nu} = \frac{\xi_\mu}{\xi} \frac{\partial}{\partial \xi} \left[ \frac{\xi^\nu}{\xi} \frac{\partial R}{\partial \xi}  \right] = \frac{\partial^2 R}{\partial \xi^2 } u_\mu u_\nu + \frac{1}{\xi} \frac{\partial R}{\partial \xi} {h}_{\mu\nu},
\end{equation*}
and contracting (\ref{einsxi}) with ${g}^{\mu\nu}$, we get the equation known in the theory of cylindric functions:
\begin{equation}
	\frac{\partial^2 R}{\partial \xi^2} + \frac{3}{\xi} \frac{\partial R}{\partial \xi} = m^2 c^4 R.
	\label{einsxitrace}
\end{equation}
The left side in this expression represents d'Alembertian of a spherically symmetric function $R(\xi)$ in the tangent 4-dimensional thermodynamic $\xi$-space. Thus, (\ref{einsxitrace}) proves to be equivalent to the Klein-Gordon equation (with negative mass-square) defined on this Killing space:
\begin{equation}
	 \left( \underset{\xi}{\square} - m^2 c^4 \right) R = 0,
	 \label{xiKG}
\end{equation}
with $\underset{\xi}{\square} = \frac{\partial^2}{\partial(\xi^0)^2} - \frac{\partial^2}{\partial(\xi^1)^2} - \frac{\partial^2}{\partial(\xi^2)^2} -\frac{\partial^2}{\partial(\xi^3)^2}$ written in Cartesian coordinates. Conversely, if, in accord with (\ref{KG}), we rewrite  (\ref{xiKG}) in pseudo-spherical coordinates in $\xi$-Minkowski space (analogue of the Milne metric in the usual space-time),
\begin{equation*}
	ds^2_\xi = d \xi^2 - \xi^2 \left( \chi^2 + \sinh^2 \chi d \Omega^2 \right),
\end{equation*}
with $\xi =  \xi_{\mu}u^\mu = \xi_0$, then we return to (\ref{einsxitrace}).

So, applying  the geometrized  kinetic $\xi$-formalism (\ref{xieins}) we have performed the transfiguration of the Einstein equations (\ref{kineins}) into the thermodynamic ``$\xi$-Gordon" equation (\ref{xiKG}). Its operator has the dimensionality of energy squared, unlike the standard Klein-Gordon operator $\left(  \square + \frac{m^2 c^2}{\hbar^2} \right)$ with the dimension of $(\text{length})^{-2}$.

The solution of (\ref{einsxitrace}) with the boundary conditions $R(\xi) = 0$ and $\partial R/\partial \xi = 0$ at $\xi \to \infty$ (zero temperature) leads to the Ricci scalar for a system under consideration as a function of $\xi$ (and, consequently, of temperature):
\begin{equation*}
	R = \text{const}  \frac{K_1(m c^2 \xi)}{\xi}.
\end{equation*}
As it should, it coincides, up to the factor $(-{\varkappa m^2 c^3})$, with the generating function $\Phi$, thereby reproducing all the chain of moments in the kinetic $\xi$-formalism.

\section{Conclusion}
The scalarization of the Raychaudhuri identity allows to develop the $\xi$-formalism employing underlying symmetries (generally speaking, homeomorphisms) in terms of the spacetime deformation tensor. This presents an additional means to study not only usual kinematical characteristics of timelike congruences but also physically relevant symmetries of metrics under consideration, as we have illustrated, in particular, for some  problems in cosmology (evolution of scalar vs. antiscalar background) and in general relativistic thermodynamics. In the latter case, the absence of horizon in antiscalar solutions represents a fundamental physical difference from the traditional black hole thermodynamics; nevertheless, the final results, if evaluated for equipotential surfaces at the horizon scale, prove to be the same. In a broader context, the transformation of kinetic Einstein's equations into the ``$\xi$-Gordon" equation in thermodynamic Killing space is in line with the general idea that the Einstein equations and thermodynamics appear to be intricately connected.

\section*{Acknowledgments} 
The work is partially funded by the program PCF BR05236322 of the Republic of Kazakhstan.

\bibliography{references}

\end{document}